\documentclass[useAMS,usenatbib,usegraphicx]{mn2e}
\usepackage{psfig}
\title[Mini-halo--star encounters]{On mini-halo encounters with stars}
\author[A.~M.~Green and S.~P.~Goodwin]{
  Anne M.~Green$^1$\thanks{anne.green@nottingham.ac.uk}
  and Simon P. Goodwin$^2$\thanks{s.goodwin@sheffield.ac.uk} \\
   $^1$ School of Physics and Astronomy, University of Nottingham,
     Nottingham, NG7 2RD, UK \\
    $^2$  Department of Physics and Astronomy, University of Sheffield,
    Sheffield, S3 7RH, UK }
\begin{document}
\date{}
\pagerange{\pageref{firstpage}--\pageref{lastpage}} \pubyear{2006}
\maketitle
\label{firstpage}
\begin{abstract} We study, analytically and numerically,
 the energy input into dark matter mini-haloes by interactions with
stars.   We find that the fractional energy input in simulations
of Plummer spheres agrees well with the impulse approximation for 
small and large impact
parameters, with a rapid transition between these two regimes.  
Using the impulse approximation the
fractional energy input at large impact parameters is fairly
independent of the mass and density profile of the mini-halo, however
low-mass mini-haloes experience a greater fractional energy input
in close encounters.  We formulate a fitting function which  encodes
these results and use it to estimate the disruption timescales of 
mini-haloes, taking
into account the stellar velocity dispersion and mass distribution. For
mini-haloes with mass $M< {\cal O}(10^{-7} M_{\odot})$ on typical
orbits which pass through the disc, we find that the estimated
disruption timescales are independent of mini-halo mass, and are of order the
age of the Milky Way. For more massive mini-haloes the estimated
disruption timescales increase rapidly with increasing mass.
\end{abstract}

\section{Introduction}

In Cold Dark Matter (CDM) cosmologies structure forms hierarchically;
small haloes form first, with larger haloes forming via mergers and
accretion. The internal structure of haloes is determined by the dynamical
processes, for instance tidal stripping and dynamical friction, which
act on the component sub-haloes.  Numerical simulations find that
substantial amounts of substructure survive within larger haloes
(Klypin et al. 1999; Moore et al. 1999) with the number density of
sub-haloes increasing with decreasing mass, down to the resolution
limit of the simulations.  Natural questions to ask are: what are the
properties of the first dark matter (DM) haloes to form in the
Universe, and do significant numbers  survive to the present day?

Weakly Interacting Massive Particles (WIMPs) are one of the best
motivated DM candidates. They generically have roughly the required
present day density, and supersymmetry (an extension of the standard
model of particle physics) provides a well-motivated WIMP candidate,
the lightest neutralino (see e.g. Bertone, Hooper \& Silk 2005).
Numerous experiments are underway attempting to detect WIMPs either
directly (via elastic scattering off target nuclei in the lab) or
indirectly (via the products of their annihilation). In both cases the
expected signals depend critically on the DM distribution on small
scales; direct detection probes the DM on sub-milli-pc (AU) scales,
while clumping may enhance the indirect signals.  Therefore the fate
of the first dark matter haloes, and the resulting dark matter
distribution on small (sub-pc) scales, is  important for practical
reasons too.

Studies of the microphysics of WIMPs show that kinetic decoupling and
free-streaming combine to produce a  cut-off in the density
perturbation spectrum for generic WIMPs  at a comoving wavenumber $k
\sim {\cal O} (1 \, {\rm pc}^{-1})$ (Hofmann, Schwarz \& St\"ocker
2001; Schwarz, Hofmann \& St\"ocker 2001; Berezinsky, Dokuchaev \&
Eroshenko 2003; Green, Hofmann \& Schwarz 2004, 2005;  Loeb \&
Zaldarriaga 2005; Berezinsky, Dokuchaev \& Eroshenko 2006) which
corresponds to a mass of order $10^{-6} M_{\odot}$.  Analytic
calculations, using linear theory and the spherical collapse model,
find that the first typical (i.e. forming from 1-$\sigma$
fluctuations) haloes form at $z\sim 60$ and have radius $R \sim {\cal
O}(0.01 \, {\rm pc})$ (Green et al.  2004, 2005).

Diemand, Moore \& Stadel (2005) carried out numerical simulations
using as input the linear power spectrum for a generic WIMP with mass
$m_{\chi}=100 \, {\rm GeV}$ (Green et al.  2004, 2005).  They used a
multi-scale technique, twice re-simulating at higher resolution an
`average' region selected from a larger simulation.  These simulations
confirmed the analytic estimates of the mass and formation red-shift
of the first mini-haloes and also provided further information about
their properties, in particular the density profiles of sample
mini-haloes. The simulations were stopped at $z\approx 26$ when the
high-resolution region began to merge with the lower resolution
surroundings, and so the subsequent evolution of the mini-haloes has
to be studied separately.

Extensive work has been done on the dynamical evolution of more
massive ($M> 10^{6} M_{\odot}$) substructure (e.g. Zentner \& Bullock
2003; Taylor \& Babul 2004; Oguri \& Lee 2004; Pe\~narrubia \&
Benson 2005).  The physics of mini-haloes is significantly different
to these more massive haloes, however. Firstly, the first generation
of mini-haloes form monolithically, rather than
hierarchically. Secondly, the amplitude of the density perturbations
on these scales is a very weak function of scale, so that haloes with
a range of masses form at the same time.  Finally, as well as being
subject to the same dynamical processes as larger sub-haloes
(e.g. tidal stripping, interactions between sub-haloes~\footnote{See
Berezinsky et al.  (2003) and (2006) for analytic studies of the
effects of interactions between mini-haloes.})  mini-haloes can lose
energy, and possibly be completely disrupted, via interactions with
compact objects such as stars.  Various authors have used the impulse
approximation to investigate the disruption of $M \sim 10^{-6}
M_{\odot}$ mini-haloes due to encounters with stars (Diemand et
al. 2005; Zhao et al. 2005a; Moore et al. 2005; Zhao et al. 2005b;
Berezinsky et al. 2006; Goerdt et al. 2006; Angus \& Zhao 2006).  The
results of these studies range from most of the mini-haloes surviving
disruption (Diemand et al. 2005; Moore et al. 2005) to most of the
mini-haloes whose orbits pass through the solar neighbourhood being
destroyed (Zhao et al. 2005a). A definitive study will have to combine
accurate calculations of the response of mini-halos to individual
interactions with simulations of mini-halo orbits in a realistic
Galactic potential, (see Moore (1993), Zhao et al. (2005b) for work in
this direction).

In this paper we use N-body simulations to investigate the accuracy of
the impulse approximation for calculating the energy input into a
mini-halo by an interaction with a star. We formulate a fitting
function which matches the results of the simulations and use it to
estimate the timescales for one-off and multiple disruption as a
function of mini-halo mass. We caution, and discuss in more detail
below, that it is actually the mass loss which is key to determining
the extent to which a mini-halo is disrupted. The relationship between
the energy input and the mass lost in an interaction is a complex, and
to some extent unresolved, problem (see e.g.  Aguilar \& White (1985)
and Goerdt et al. (2006)).

\section{Previous calculations}

The duration of a typical star--mini-halo encounter is far shorter
than the dynamical time scale of the mini-halo, therefore the impulse
approximation holds and the interaction can be treated as
instantaneous (Spitzer 1958). More specifically, the validity of the
impulse approximation can be parameterised by the adiabatic parameter
(Gnedin \& Ostriker 1999) $x=\omega \tau$ where $\omega= \sigma(b) /b$
is the orbital frequency of particles at a distance $b$ from the
centre of the mini-halo and $\tau=2R/v$ is the duration of the
encounter.  The impulse approximation is valid if $x \ll 1$, or
equivalently $R/b \ll v/\sigma(b)$. As the typical relative velocity
of encounters 
($v \sim{\cal O} (10-100 \, {\rm km \, s}^{-1})$) is far larger than
the mini-halo velocity dispersion ($\sigma(R) \sim {\cal O}  (1 \,
{\rm m \, s}^{-1})$), then only for very rare, slow interactions at very
small impact parameters will the  impulse approximation be
violated. The change in the  velocity of a particle within an extended
body of radius $R$ at  position ${\bf r}$ relative to the centre of
the body due to  an impulsive interaction with a perturber of mass
$M_{\star}$, moving  with relative velocity ${\bf v}$ at an impact
parameter ${\bf b}$,  such that $b \gg R$, is given by (Spitzer 1958)
\begin{equation}
\delta {\bf v} \approx \frac{2 G M_{\star}}{v b^2} \left[ 2 ({\bf
r}.{\bf e_{\rm b}}) {\bf e}_{\rm b} + ({\bf r}.{\bf e}_{\rm v}) {\bf
e}_{\rm v} - {\bf r} \right]  \,,
\end{equation}
where ${\bf e}_{\rm v}$ and ${\bf e}_{\rm b}$ are unit vectors
perpendicular to ${\bf v}$ and ${\bf b}$ respectively.  The energy
input, per unit mass, for an individual particle is  $\delta E = {\bf
v} . (\delta {\bf v}) + 0.5 (\delta {\bf v})^2$, and the total energy
input into the body is then found by integrating over the density
distribution. For a spherically symmetric body the first term averages
to zero and, using the approximation that  $({\bf r}.{\bf e}_{\rm
v})^2= ({\bf r}.{\bf e}_{\rm b})^2 \approx r^2/3$, the total energy
input is given by
\begin{equation}
\label{deltaEa}
\Delta E(b)   \approx  \frac{ 4 \alpha^2}{3}  \frac{ G^2 M_{\star}^2 M
   R^2}{v^2 b^4} \,,
\end{equation}
where
\begin{equation}
\label{a}
\alpha^2 = \frac{<r^2>}{R^2} \equiv \frac{1}{R^2}  \left[
     \frac{\int_{0}^{R} {\rm d}^3 {\bf r} \, r^2 \rho(r)} {M} \right]
     \,,   \\
\end{equation}
is the root mean square radius.

For small impact parameter interactions, $b/R \rightarrow 0$,
(e.g. Gerhard \& Fall 1983, Carr \& Sakellariadou 1999)
\begin{equation}
\delta {\bf v} \approx \frac{2 G M_{\star}}{v} \left[ \frac{({\bf
r}.{\bf e}_{\rm v})  {\bf e}_{\rm v} - {\bf r}}{{\bf r}^2 - ({\bf
r}.{\bf e}_{\rm v})^2} \right]  \,,
\end{equation}
so that the energy input is given by
\begin{equation}
\label{deltaEb}
\Delta E(b=0)   \approx 3 \beta^{2} \frac{G^2 M_{\star}^2 M}{v^2 R^2}
\,,
\end{equation}
where
\begin {equation}
\label{b}
\beta^2  =   <r^{-2}> R^2  \equiv R^2  \left[ \frac{\int_{0}^{R} {\rm
     d}^3 {\bf r} \, r^{-2} \rho(r)} {M} \right] \,,
\end{equation}
is the root mean square inverse radius. Carr \& Sakellariadou (1999)
(drawing on Gerhard \& Fall (1983)) interpolate between the  $b \gg R$
and $b \ll R$ regimes using
\begin{eqnarray}
\delta {\bf v} & = & \frac{2 G M_{\star}}{v} \frac{1}{b^2+ (2 r^2/3)}
   \nonumber \\  & \times & \left[ \frac{2 b^2}{b^2 + (2 r^2/3)} ({\bf
   r}.{\bf e_{\rm b}}) {\bf e}_{\rm b} + ({\bf r}.{\bf e}_{\rm v})
   {\bf e}_{\rm v} - {\bf r} \right]  \,,
\end{eqnarray}
so that
\begin{eqnarray}
\label{deltaE1}
&&  \Delta E(b) \approx \frac{4}{3} \left( \frac{ G M_{\star}}{v b^2}
 \right)^2 \nonumber \\ && \times \int_{0}^{R} {\rm d}^3 {\bf r} \,
 r^2 \rho(r) \left( 1 + \frac{ 4 r^4}{9 b^4} \right)  \left( 1 +
 \frac{2 r^2}{3 b^2} \right)^{-4} \,.
\end{eqnarray}

\begin{table*}
\begin{center}
\begin{tabular}{|c|c|l|l|l|l|l}
\hline halo & profile & $\alpha^2 R^2 \, ({\rm pc}^2)$  & $\beta^2/R^2
\, ({\rm pc}^{-2})$ &  E/M \, (erg/$M_{\odot}$)  & M \, ($M_{\odot}$)  &
$r_{{\rm p/c/s}}$ \, (pc) \\ 
\hline 
1 & plummer & $2.6 \times 10^{-4}$ & $1.5 \times 10^{4}$ & $-1.5
\times 10^{38}$ &$1.0 \times 10^{-4}$ & 0.013\\ 
1 & CIS  & $3.2 \times 10^{-4}$ & $3.2 \times 10^{4}$  &
$-7.2 \times 10^{37}$ & $6.0 \times 10^{-5}$ & 0.0032 \\ 
1 & NFW  & $3.2 \times 10^{-4}$ &
-& $-7.8 \times 10^{37}$ & $6.7 \times 10^{-5}$ & 0.0014\\ 
\hline 
2 & plummer
& $3.2 \times 10^{-5}$ &  $7.0 \times 10^{4}$  & $-2.1 \times 10^{37}$ &$2.1 \times 10^{-6}$ &  0.0092\\ 
2 &
CIS  & $2.7 \times 10^{-5}$ & $1.2 \times 10^{5}$ & $-1.1 \times 10^{37}$ & $1.6 \times 10^{-6}$ &  0.0032  \\ 
2
& NFW  & $2.7 \times 10^{-5}$ &- & $-5.1 \times 10^{37}$ &$1.4 \times 10^{-6}$ & 0.011    \\
\hline 
3 & plummer &  $3.0 \times 10^{-5}$ & $8.1 \times 10^{4}$ & $-1.0 \times 10^{37}$ &$1.3 \times
10^{-6}$ & 0.0076\\ 
3 & CIS  & $2.4 \times 10^{-5}$ & $2.0 \times 10^{5}$ & $-1.0 \times 10^{37}$ & $1.2
\times 10^{-6}$ & 0.0019 \\ 
3 & NFW  & $2.6 \times 10^{-5}$ &- & $-5.2 \times 10^{37}$ & $
1.4 \times 10^{-6}$ & 0.0076\\ \hline
\end{tabular}
\end{center}
\caption[bf]{Structure parameters, binding energy per unit mass 
and mass of  the
best-fit profiles for the 3 sample haloes (using only the data at
radii greater than the force softening). For the energy and mass
calculations a sharp cut-off is taken at $R=r_{200}(z=26)= 0.03 \,
{\rm pc}$ and $0.008 \, {\rm pc}$ for  halo 1 and haloes  2 $\&$ 3
respectively. The final column is the Plummer, core or scale  radius
for the Plummer, CIS and NFW profiles
respectively.}
\end{table*}

Moore (1993) simulated encounters between globular clusters, modelled
with King profiles, and massive ($\sim 10^{4} M_{\odot}$) black holes.
He found that the energy input was well fitted by
\begin{equation}
\label{deltaEapprox}
\Delta E(b)= \frac{\Delta E(b=0)}{\left[ 1 + (b/R) \right]^4} \,.
\end{equation}
This fitting function will, however, only reproduce the asymptotic
limits, equations~ (\ref{deltaEa}) and (\ref{deltaEb}), if $\alpha^2
\approx 9 \beta^2/ 4$, which need not be (and we will see is not) the
case in general. A simple modification to equation~(\ref{deltaEapprox})
\begin{equation}
\label{deltaEapprox2}
\Delta E(b)= \frac{\Delta E(b=0)}{\left[ 1 +  (b A^{-1/4}/R)
\right]^4} \,,
\end{equation}
with $A = 4 \alpha^2 / 9 \beta^2$, produces a function with the
correct asymptotic limits in general.

For later convenience we write the fractional energy input, which  is
simply $\Delta E(b)$ divided by the total energy of the mini-halo 
$E= \gamma G M^2/R$, where $\gamma$ is a constant of order one which
depends on the density profile (and cut-off radius if one is imposed),
as
\begin{equation}
\frac{\Delta E(b)}{E} = \left(\frac{\Delta E(b)}{E} \right)_{\rm fid}
                  \left[ \left( \frac{M_{\star}}{M_{\sun}}  \right)
                  \left(\frac{300 \,{\rm km \, s^{-1}}}{ v} \right) \
                  \right]^2 \,,
\end{equation}
where $(\Delta E(b)/{E})_{\rm fid}$ is the fractional energy input in an
interaction with a fiducial star with mass $M_{\star}=1 M_{\odot}$ and
relative velocity $v=300 \, {\rm km \, s^{-1}}$.

\section{Application to mini-haloes}

\subsection{Density profiles}

We use three benchmark density profiles:
\begin{enumerate}
\item {{\em Plummer sphere}
\begin{equation}
\rho(r) = \frac{\rho_{0}}{\left[ 1 + (r/r_{\rm p})^2 \right]^{5/2}} \,,
\end{equation}
This profile (Plummer 1915), which has a central core and asymptotes
to $r^{-5}$ at large radii, is commonly used to model star clusters.
It is not a good fit to simulated CDM halos or subhalos, however it is
a convenient choice for testing the impulse approximation against
numerical simulations as it has a simple form for the density and
velocity distributions (see Aarseth, H\'enon \& Weilan 1974). In
addition, the rapid fall off of the density at large radii means that
there are no subtleties involved in imposing a truncation radius (see
e.g. Kazantzidis, Magorrian \& Moore 2004), and the Plummer sphere is
also  stable when isolated.  }

\item{{\em cored isothermal sphere (CIS)}
\begin{equation}
\label{rhoISSc}
\rho(r)= \rho_{0} \frac{r^2 + 3 r_{\rm c}^2} {\left( r^2+ r_{\rm c}^2
     \right)^2} \,,
\end{equation}
The cored isothermal sphere is a better (although still not good)
approximation to the mini-haloes, is amenable to analytic
calculations and allows us to investigate the impact of a central core
and more gradual fall-off at large radii to the fractional energy input.  }

\item{{\em Navarro, Frenk, White (NFW)}
\begin{equation}
\rho(r)=\frac{\rho_{0}}{(r/r_{\rm s})[1+(r/r_{\rm s})]^2} \,,
\end{equation}
The Navarro, Frenk, White profile (Navarro, Frenk \& White 1996, 1997)
fits the density distribution, outside the very central regions,  of
simulated galactic scale and larger dark matter haloes well and  is
often used to model massive dark matter haloes. However, mini-haloes
form monolithically, rather than by hierarchical mergers like
`standard' dark matter haloes, and it is not clear that they will have
the same density profile. The NFW profile does, however,  provide a
reasonably good fit to the density profiles of the  mini-haloes from
Diemand et al.'s simulations.  }

\end{enumerate}

We find the best fit for each of these profiles for the three typical
haloes in Fig.~2 of Diemand et al. (2005) using only the data points
(density averaged within radial bins) at radii greater than the force
resolution (using all the data points does not significantly change
the best fit parameters).  We refer to the haloes denoted by squares,
stars and circles in their figure as haloes 1, 2 and 3 respectively.

The CIS and NFW profiles have infinite mass and energy unless a
cut-off radius is imposed by hand. For definiteness, and to allow
comparison with previous work on mini-halo disruption, we use the
radius at which the halo density is 200 times the cosmic mean density
at $z=26$ when Diemand et al.'s simulations are stopped and the sample
haloes studied.  For halo 1 $r_{200}(z=26)=0.03 \, {\rm pc}$ while for
haloes 2 and 3 $r_{200}(z=26)=0.008 \, {\rm pc}$.  For the Plummer
profile a cut-off is not in principle needed, however we use the same
values for the radii for consistency.

In Table~1 we give the values of the Plummer, core and scale radii
(as appropriate), the mass ($M$), initial energy per unit mass ($E/M$)
and the structure parameters, $\alpha^2 R^2$ and $\beta^2/R^2$,  Our
values of the structure parameters are slightly different to those of
Carr \& Sakellariadou (1999) as they define the cluster radii
differently. Haloes 2 and 3 have roughly the same mass $\sim 10^{-6}
M_{\odot}$ depending at the tens of per-cent level on the profile
used.  Halo 1 is a factor of $\sim 50$ more massive~\footnote{It
appears that the $5.1 \times 10^{-6} M_{\odot}$ for halo one in the
caption of fig.~2 of Diemand et al. (2005) is a typo and should be
$5.1 \times 10^{-5} M_{\odot}$.}.  We calculate the total
energy by calculating the potential, and hence the velocity dispersion
and kinetic and potential energy densities, from the density
profiles. The truncation at finite radii means that the resulting
haloes are not in virial equilibrium.  For all three profiles for halo
1 and the CIS and NFW profiles for haloes 2 and 3 the deviation is
relatively small. The best fit Plummer spheres for haloes 2 and 3 have
$r_{\rm s} \sim {\cal O} (r_{200}(z=26))$ and the resulting systems
are far from virial equilibrium.

The values of $\alpha^2$, which parameterises the energy input for
large impact parameter encounters, only vary by a factor of $\sim 2$
between different haloes and density profiles reflecting
the fact that the sample halos have similar mean densities.  
However $\beta^2$,
which parameterises the energy input in the $b \rightarrow 0$ limit
varies significantly and is in fact infinite for the NFW profile. It
can be seen from the definition of $\beta^2$, equation~(\ref{b}), that
$\beta^2$ is formally infinite  for any profile with a central cusp
$\rho(r) \sim r^{-\gamma}$ with $\gamma \geq 1$. The WIMP density can
not in fact become arbitrarily high in the central regions of a
mini-halo; if the density becomes sufficiently high the WIMPs will
annihilate, reducing the density to some maximum value $\rho_{\rm
max}$ so that the halo has a (small) core: $\rho(r)= \rho_{\rm max}$
for $r< r_{\rm core}$.  The density and size of the core can be
estimated by calculating the density for which the annihilation time
scale is less than the Hubble time (c.f. Berezinsky, Gurevich \& Zybin
1992) \begin{equation} \frac{\rho_{\rm max} < \sigma_{\chi \chi} v >
}{2 m_{\chi}} < \frac{1}{ 10^{10} {\rm yr}} \,.
\end{equation}
Using `typical' values  for the WIMP mass and velocity averaged
cross-section, $m_{\chi} = 100 \, {\rm GeV},  < \sigma_{\chi \chi} v >
= 3 \times 10^{-32} \, {\rm m}^{3} {\rm s}^{-1}$, we find $\rho_{\rm
max} = 4 \times 10^{-13} \, {\rm kg \, m}^{-3} = 4 \times 10^{13}
\rho_{\rm c}(z=0)$. For the best fit NFW profiles  $r_{\rm core} \sim
10^{-10} {\rm pc}$. Taking this effect into  account leads to finite
values for $\beta^2$, but they are still large ($\sim 10^{7}$). The
energy input only reaches its asymptotic value, however, for tiny
($\sim r_{\rm core}$), and hence extremely rare, impact parameters.

\subsection{Simulations of star -- mini-halo encounters}

We use the {\sc dragon} smooth particle hydrodynamics code
(e.g. Goodwin, Whitworth \& Ward-Thompson 2004a,b; Hubber, Goodwin \&
Whitworth 2006) with hydrodynamics turned-off as an $N$-body code.
{\sc dragon} uses a Barnes-Hut (1986)-type tree and we set the opening
angle to be small to increase the accuracy of the force calculations
between DM particles.  The forces between DM particles and the star
are all computed by direct summation. This physical situation,
interaction of an extended body with a far more massive compact
object, has, to our knowledge, not been studied numerically before and
we carried out extensive testing to ensure the reliability of the
results.  In particular, the masses of the DM particles are a factor
$\sim {\cal O} (10^{9})$ less massive than the perturbing star,
requiring numerical care to be taken.

We generate the initial conditions for the Plummer sphere mini-haloes
using the prescription of Aarseth et al. (1974), assuming that the
haloes are initially in virial equilibrium. Left isolated, the
mini-haloes remain in equilibrium, and the energy conservation of the
code is $\sim~10^{-5}$ over timescales far in excess of a typical
mini-halo--star interaction timescale  ($\sim 50$~kyr). A star of mass
$M_{\star}$ is then placed 1$\,{\rm pc}$ away from the halo
approaching  it at velocity $v$, with an impact parameter $b$.

We conduct simulations with $N=5000$ DM particles with a Plummer force
softening between DM particles of $\epsilon = 10^{-3}$~pc.  The forces
due to the star are softened with a significantly smaller softening
length of $10^{-4}$~pc.  The softening between DM particles is rather
large, but we wish to subdue any 2-body interactions between DM
particles.  Tests conducted with $\epsilon = 10^{-4}$ and $10^{-2}$~pc
show {\em no} difference in the results.  Similarly, increasing the
particle numbers to $N=10000$ and $20000$ we found no significant (or
systematic) changes.  This convergence is not surprising as the energy
input is entirely due to the encounter with the star whose force is
accurately calculated with a low softening length, and we are only
concerned with the energy input to the halo, and not in the details
of relaxation and/or mass loss after the impulse has occurred (which
will involve interactions between the halo particles and may require
a larger number particles for convergence e.g. Goerdt et al. 2006).

We ran a large ensemble of simulations covering a wide range of
$M_{\star}-v-b$ parameter space: 
$0.215 < M_{\star}/M_\odot < 30$, $1 < v/({1 \, \rm
km\,\,\,s}^{-1}) < 400$~\footnote{Although interactions with relative
speeds at the lower end of this range are extremely rare we consider
them in order to test the validity of the impulse approximation.} and
$-5 < {\rm log}_{10}(b/1 \, {\rm pc}) < 1$.  With  $N=5000$, each
simulation took an average of 20 minutes on a desktop PC.

\subsection{Fractional energy input}
\label{fel}

\begin{figure}
\includegraphics[width=\linewidth]{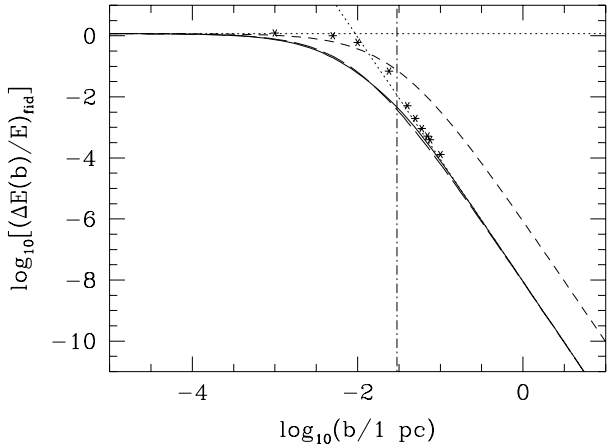}
\caption{The fractional energy input, $(\Delta E(b)/{E})_{\rm fid}$,
in an interaction with a fiducial star with mass $M_{\star}=1
M_{\odot}$ and relative speed $v=300 \, {\rm km \, s^{-1}}$ for the
best fit Plummer profile for halo 1.  Solid line from numerically
integrating equation~(\ref{deltaE1}), dotted lines asymptotic limits,
equations~(\ref{deltaEa}) and (\ref{deltaEb}),  short dashed line
using the original  fitting function, equation~(\ref{deltaEapprox}),
and long dashed line the modified fitting function,
equation~(\ref{deltaEapprox2}).  Stars from numerical simulations.
The dot-dashed line is the radius of the mini-halo.  }
\end{figure}

\begin{figure}
\setlength{\unitlength}{\linewidth}
\includegraphics[width=6.0cm,angle=270]{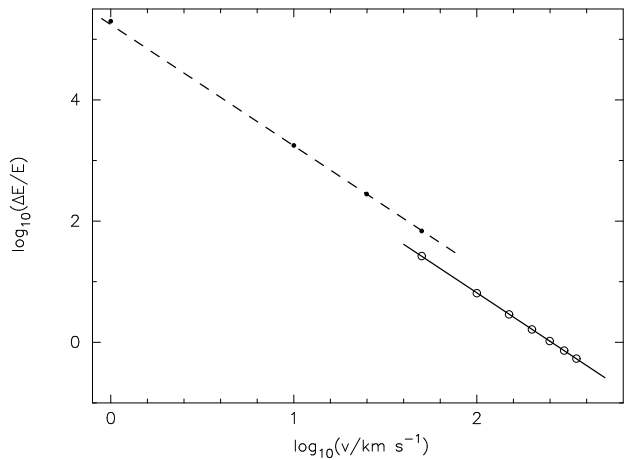}
\caption{The fractional energy input, $\Delta E/E$, from simulations
of the
best fit Plummer profile for halo 1  as a function of relative velocity
for encounters with a perturber of mass $M_{\star} = 1 M_\odot$.  The
open circles have an impact parameter of $b=10^{-2}$~pc, while the filled
circles have $b=10^{-5}$~pc. The fractional energy input scales as $v^{-2}$ as 
expected (lines of gradient $-2$ have been added to
aid the eye).}
\label{fig:v2}
\end{figure}

\begin{figure}
\setlength{\unitlength}{\linewidth}
\includegraphics[width=6.0cm,angle=270]{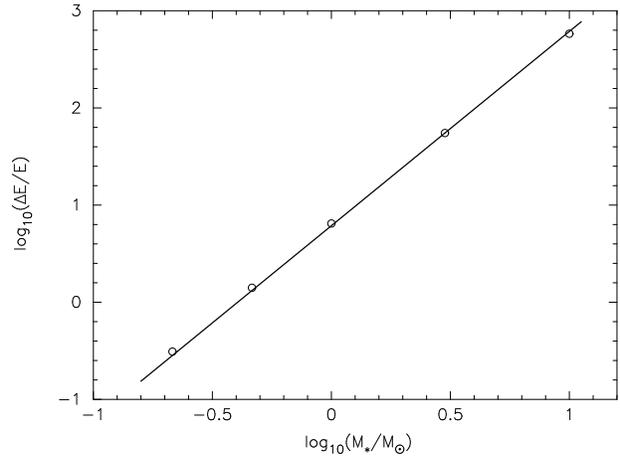}
\caption{The fractional energy input, $\Delta E/E$, from simulations of the
best fit Plummer profile for halo 1
 as a function of perturber mass for encounters with
a relative velocity $v=100$~km s$^{-1}$ at an impact parameter of 
$b=10^{-2}$~pc. The fractional energy input scales as $M_{\star}^{2}$ 
as expected (a line of gradient $2$ has been added to aid the eye).}
\label{fig:m2}
\end{figure}

The fractional energy input, $(\Delta E(b)/{E})_{\rm fid}$, in an
interaction with a fiducial star with mass $M_{\star}=1 M_{\odot}$ and
relative velocity $v=300 \, {\rm km \, s^{-1}}$ is plotted in Fig.~1
for the best fit plummer sphere for halo 1. This fiducial velocity was
chosen as an isothermal sphere with circular velocity $v_{\rm c}=220
\, {\rm km \, s}^{-1}$ (i.e. representing the Milky Way) has root mean
square speed of 270 ${\rm km \, s}^{-1}$. In reality, interactions
will have a range of velocities and perturber masses.  In
Figs.~\ref{fig:v2} and~\ref{fig:m2} we plot the fractional 
energy input as a function
of relative velocity and perturber mass, showing that it scales as $v^{-2}$ and
$M_{\star}^2$ respectively as expected from equation~(\ref{deltaEa}).  In
Fig.~\ref{fig:v2} we also show that the $v^{-2}$ proportionality is
independent of the impact parameter and holds down to very small
({\cal O}($1$~km~s$^{-1}$)) relative velocities.

For a given perturber mass and relative velocity, we see from  Fig.~1
that the large and small $b$ asymptotic limits are in excellent
agreement with the full analytic calculation using
equation~(\ref{deltaE1}). As expected, the original fitting function
significantly over-estimates the energy input at large $b$.  The
modified fitting function, designed to reproduce the asymptotic
limits, matches well the calculation using equation~(\ref{deltaE1})
for all $b$.  In the simulations, however, the transition between the $b \ll R$
and $b \gg R$ regimes happens very rapidly and the energy input is
well approximated, for all $b$, by the minimum of the asymptotic
limits:
\begin{equation}
\label{deltaEapprox3} \Delta E(b) = \frac{G^2 M_{\star}^2 M}{v^2}
      \times {\rm min} \left( \frac{ 4 \alpha^2 R^2}{3 b^4}  , \frac{3
         \beta^{2}}{R^2}    \right) \,.
\end{equation}
In the $b \sim R$
regime the energy input in the simulations is significantly larger
than that from the analytic impulse approximation calculation. This
may be indicating that in this regime, due to the assymetry of the
interaction, the $(\delta {\bf v}). {\bf v}$ term in the total energy
input does not average to zero. It would be interesting to examine 
whether the energy input for $ b \sim R$ depends on the
mini-halo density profile.

In Fig.~4 we plot the fractional energy input from an interaction with a
fiducial star with mass $M_{\star}=1 M_{\odot}$ and relative speed $v= 300 \,
{\rm km \, s}^{-1}$ for the the best fit profiles for all 3 haloes
calculated using equation~(\ref{deltaE1}).  The fractional energy input for
close interactions, which is proportional to $ \beta^2 M / E R$,
varies by a factor of $\sim 3 $ for a given halo and is $\sim 100 $
times larger for the lighter haloes 2 and 3. This indicates that
smaller, lighter mini-haloes are far more susceptible to disruption by
close encounters. For large impact parameter interactions ($b \gg R$)
the fractional energy input, which is proportional to $ \alpha^2 M R^2
/ E$ varies only weakly (by a factor of $\sim 3$) between haloes and
profiles, with the spread in values for different profiles for a given
halo being comparable to that for different haloes for a fixed
profile. 

\begin{figure}
\setlength{\unitlength}{\linewidth}
\includegraphics[width=\linewidth]{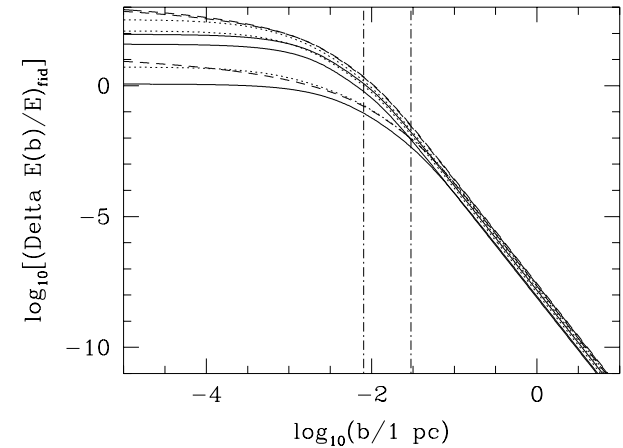}
\caption{The fractional  energy input in an interaction
with a fiducial star with mass $M_{\star}=1 M_{\odot}$ and relative speed
$v= 300 \, {\rm km \, s}^{-1}$, $(\Delta E(b)/E)_{\rm fid}$, calculated
using equation~(\ref{deltaE1})
for the best fit Plummer (solid), CIS (dotted) and NFW (dashed)
profiles for (from bottom to top) haloes 1, 2 and 3. The dot-dashed
lines are the radii of the mini-haloes (0.03~pc for halo 1, and 
0.008~pc for haloes 2/3).}
\end{figure}

This behaviour can be qualitatively understood by considering the
asymptotic fractional energy input for a uniform density sphere (with
$\rho=\rho_{0}$ for $r< R$ and $\rho=0$ otherwise):
\begin{displaymath}
\frac{\Delta E(b)}{E}   \propto 
  \left\{ \begin{array}{ll}
  \frac{1}{\rho_{0}} \hspace{1.3cm} 
   (b \gg R)  \,, \\ \nonumber
   \frac{1}{\rho_{0} R^4}
      \hspace{1.0cm} (b \ll R) \,.
\end{array} \right.
\end{displaymath}
On galactic scales the red-shift at which a given scale goes non-linear,
and hence the characteristic density of typical haloes, is strongly scale
dependent. The comoving scales corresponding to the 
mini-haloes ($k > {\cal O}(0.1
\, {\rm pc}^{-1})$) entered the horizon during the radiation dominated
epoch, where (CDM) density perturbations grow only logarithmically.
The size of the density perturbations, at fixed red-shift, on these scales
is therefore only logarithmically dependent on the comoving
wavenumber. Consequently the red-shift at which a given physical scale
goes non-linear, and hence the characteristic density of the resulting
haloes, is only weakly (roughly logarithmically) dependent on the scale
(see e.g.  Green et al. 2005).  Neglecting this weak scale dependence
and making the approximation that $\rho_{0} \sim {\rm const}$, then
$\Delta E/E \sim {\rm const}$ for $b \gg R$ and $\Delta E/E \sim
M^{-4/3}$ for $b \ll R$. These scalings are in broad agreement with
the trends found for the three sample haloes. The weak scale dependence
of the red-shift of non-linearity will lead to more massive haloes
typically
having lower characteristic densities and hence being slightly more
susceptible to disruption. This scale dependence is relatively small
however and is comparable in magnitude to the dependence on the
mini-halo density profile.

This behaviour, along with the results from the numerical
simulations for the Plummer sphere, indicates that a reasonable 
approximation to the fractional energy input is
given by a sudden transition between the asymptotic $b<</>> R$ regimes:
\begin{displaymath}
\label{deltaEapprox4}
\left(\frac{\Delta E}{E} \right)_{\rm fid} = \left\{
  \begin{array}{ll}
    \left(\frac{\Delta E}{E} \right)_{{\rm fid, s}} 
      \left( \frac{1 {\rm pc}}{b} \right)^{4} \hspace{1.0cm} b > b_{\rm s} \,,
    \nonumber \\
  \left(\frac{\Delta E}{E} \right)_{{\rm fid}, 0} 
      \left( \frac{1 {\rm pc}}{b_{\rm s}} \right)^{4} 
\hspace{1.0cm} b < b_{\rm s} \,,
\end{array} \right.
\end{displaymath}
where
\begin{equation}
\left(\frac{\Delta E}{E} \right)_{{\rm fid, s}}  = 
    \frac{ 4 \alpha^2}{3} 
   \frac{ G^2 M_{\odot}^2  M R^2}{(300 \, {\rm km
 \, s}^{-1})^2 (1 \, {\rm pc})^4} \approx 1 \times 10^{-8} \,,
\label{sudden}
\end{equation}
is the asymptotic large $b$ slope
and the transition between the two regimes occurs, for the 3 sample haloes,
at
\begin{equation}
\label{ps}
b_{\rm s} = \left( \frac{4 \alpha^2}{9 \beta^2} \right)^{1/4} R =A^{1/4} R
        \approx
       (0.3-0.45) R \,.
\end{equation}
The large variation in the value of $\beta^2$ for the different 
haloes/profiles has a relatively small (less than a factor of 2)
effect on the value of $A$, however taking  
this R/M dependence into account is crucial for obtaining the correct
$b\ll R$ asymptotic behaviour.
The NFW profile, however, is slightly problematic. As discussed 
above, its (very large) asymptotic value of $\beta^2$ is only reached
for tiny impact parameters. A reasonable prescription for this profile
is to use the asymptotic fractional energy input at $b=0.1R$ to calculate the 
value of $\beta^2$.

\vspace{0.5cm}

It should be emphasised that the representative sample haloes studied
in detail by Diemand et al. presumably form from `typical', $\sim 1-2 \sigma$,
fluctuations. Similarly, our discussion (above) of the scaling of the
energy input with the halo mass implicitly assumed that the haloes form
from similar sized over-densities. Mini-haloes which form from rarer large
over-densities will be denser and hence more resilient to disruption
(Berezinsky et al. 2003, 2006; Green et al. 2004, 2005). More
specifically, in the spherical collapse model, a halo forming on a given
comoving scale from an $N \sigma$ fluctuation will have $R \propto
1/N$, $M \sim {\rm constant}$, and characteristic density $\rho \propto
N^3$ (Green et al. 2005). We therefore expect that the fractional
energy input in close encounters will be far smaller for haloes formed
from rarer, larger, fluctuations. The quantitative effect on the fractional
energy input will depend on exactly how the characteristic density and 
density profile
scale with the size of the overdensity from which the mini-halo forms.

\subsection{One-off disruption}

We now use the criterion $\Delta E(b_{\rm c})/E=1$ and the sudden
transition approximation developed in Section~\ref{fel} above, to
estimate the critical impact parameter $b_{\rm c}$, below which the
energy input in a single encounter is larger than the binding
energy. Taken at face value, an energy input $\Delta E (b)/E > 1$ might
appear to imply that the mini-halo is completely disrupted.  In
reality, however, the reaction of a system to a sudden change of
energy, and in particular the relationship between the energy input
and the mass lost, is non-trivial (see e.g.  Aguilar \& White 1985;
Goodwin 1997; Gnedin \& Ostriker 1999; and, for the specific case of
mini-halo interactions with stars, Goerdt et al. 2006).  The system
will expand and attempt to revirialise, and during this process
two-body encounters will redistribute energy between particles. The
simple criterion $\Delta E (b_{\rm c})/E = 1$ allows us to make an
estimate of the impact parameter below which a mini-halo will lose a
substantial fraction of its mass in a single encounter (which for
compactness we refer to as `one-off disruption').  A detailed
calculation of the mass loss, however, requires numerical simulations
of the revirilisation and energy redistribution processes
(c.f. Goerdt et al. 2006).

One-off disruption can not occur if the asymptotic fractional energy 
input as $b$ tends to zero is less than one. This is the case if
\begin{equation} 
\label{pcritapprox1}
\left( \frac{M_{\star}}{M_{\odot}} \frac{300 \, {\rm km \, s}^{-1}}{v}
   \right)  < 
    \left(\frac{\Delta E}{E} \right)_{{\rm fid}, s}^{-1/2} 
      \left( \frac{b_{\rm s}}{1 {\rm pc}} \right)^2  \,.
\end{equation}
Otherwise, 
\begin{equation}
\label{pcritapprox2}
 \frac{b_{\rm c}}{1 \, {\rm pc}}  =  
   \left(\frac{\Delta E}{E} \right)_{{\rm fid}, s}^{1/4} 
     \left( \frac{M_{\star}}{M_{\odot}}  
  \frac{300 \, {\rm km \, s}^{-1}}{v} \right)^{1/2} \,.
\end{equation}

\begin{figure}
\setlength{\unitlength}{\linewidth}
\includegraphics[width=6.0cm,angle=270]{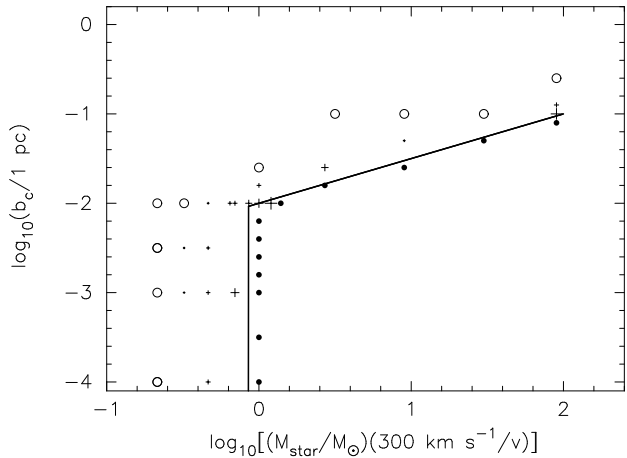}
\caption{The maximum impact parameter for which one-off disruption
can occur, $b_{\rm c}$, for the best fit plummer profile for halo 
1 (which has $R=0.03$ pc and $r_{\rm p}=0.013$ pc). 
The solid line shows the analytic calculation using 
the sudden transition approximation, 
equation~(\ref{pcritapprox2}). The symbols are the results
of numerical simulations: 
open circles for $(\Delta E(b)/E)<0.05$, filled circles for 
$(\Delta E(b)/E)>1$ (potential one-off disruption) and the size of crosses is
proportional to $(\Delta E(b)/E)$ in the intermediate regime.}
\end{figure}

In Fig.~5 we plot $b_{\rm c}$ as a function of $M_{\star}/v$ for the
best fit Plummer profile for halo 1 calculated using the analytic
expressions derived from the sudden transition approximation
(eqns.~\ref{pcritapprox1}-\ref{pcritapprox2}). We also plot the energy
input in numerical simulations, demonstrating that the sudden transition
approximation
provides a good fit to the transition between the $(\Delta E(b)/E)>1$
and $(\Delta E(b)/E)< 1$ regimes. The critical impact parameter is
quite sensitive to the properties of the perturbing star. Thus a
full calculation of mini-halo disruption will have to take into
account the stellar mass and velocity distributions.

\section{Implications and open issues}

\subsection{Disruption timescales}

As discussed in the introduction (see also Zhao et al. 2005b; Goerdt
et al. 2006; Angus \& Zhao 2006) an accurate calculation of the
mini-halo survival probability distribution will require the 
combination of simulations of mass loss with orbits in a realistic Galactic
potential. In this section though, we use the results of our energy
input studies in Section 3, in particular the sudden transition
approximation, to estimate the disruption timescales for typical
mini-haloes as a function of mass for some benchmark orbits.

\subsubsection{One-off disruption}

For the simplified situation where all perturbers have the same velocity 
and mass then the rate at which encounters with
impact parameter smaller than $b_{\rm c}$, the critical value for which
the energy input is larger than the binding energy, occur is
\begin{equation}
\frac{{\rm d} N}{{\rm d} t} = \pi n v b_{\rm c}^2 \,,
\end{equation}
where $n$ is the perturber number density.
Taking the stellar mass and relative speed 
to be fixed at $M_{\star}=0.5 M_{\odot}$ and $v= 270 \, {\rm km \, s}^{-1}$,
halo 1 will never undergo one-off disruption, while for haloes 2 and 3, using
the sudden transition approximation, the
critical impact parameter for one-off disruption is $0.0075 \, {\rm pc}$. 
Taking a disc mass density of 
$0.023 \, M_{\odot} \, {\rm pc}^{-3}$,~\footnote{This corresponds
to a surface density of $46 \, M_{\odot} \, {\rm pc}^{-2}$ (Kuijken \& Gilmore
1989) over a height of $2\, {\rm kpc}$.} we find a one-off disruption 
timescale, $t_{\rm dis} \approx 1/({\rm d} N/{\rm d} t)$,
for halo 2/3  of $0.5  \, {\rm Gyr}$ for (rare) halo 
orbits which lie entirely
within the Galactic disc for $b_{\rm c}= 0.0075 \, {\rm pc}$.
This indicates that a $10^{-6} M_{\odot}$ mini-halo which spends most of
its time in the disc will undergo a change in its
energy which is large compared
to its binding energy and hence lose a significant fraction of its mass.
The density of stars in the spheroid
is significantly smaller, $\sim 10^{-5} M_{\odot} \, {\rm pc}^{-3}$, and
declines rapidly with increasing Galactocentric radius, therefore mini-haloes
on orbits which never pass through the disk are extremely unlikely
to experience a close encounter which removes most of their mass.
Most mini-haloes will however be on 
intermediate orbits and spend some fraction of their time passing 
through the disc. For instance a mini-halo on a circular polar orbit
 at the solar radius with speed $v=270 \, {\rm km \, s}^{-1}$
would spend a fraction $\sim 0.08$ of its time within the disc, giving 
a disruption timescale of $6 \, {\rm Gyr}$. Therefore, in the
inner regions of the MW where orbits pass through the disc,
the timescale on which $10^{-6} M_{\odot}$ haloes which
experience significant mass loss in a single interaction
is of order the age of the 
Milky Way and a more sophisticated calculation is required.

Generalising to the more realistic case of 
a population of perturbers with a range of speeds
and masses
the rate at which interactions with impact
parameters smaller than the critical impact parameter for
potential one-off disruption to occur becomes
\begin{equation}
\frac{{\rm d} N}{{\rm d} t} = 
\int \int \frac{{\rm d}^2 n}{{\rm d} M_{\star} {\rm d} v} 
   \pi v b_{\rm c}^2(M_{\star}/v) {\rm d} M_{\star} {\rm d} v \,,
\end{equation}
where ${\rm d}^2 n/ {\rm d} M_{\star} {\rm d} v$ is the number
density of stars with mass between $M_{\star}$ and $M_{\star} + {\rm d} 
M_{\star}$ and relative speed between $v$ and $v + {\rm d} v$. We assume
that that mass and speed distributions are independent so that
\begin{equation}
\frac{{\rm d}^2 n}{{\rm d} M_{\star} {\rm d}v} = \frac{{\rm d} n_{1}}
 {{\rm d} M_{\star}} \frac{{\rm d} n_{2}} {{\rm d} v} \,. 
\end{equation}
For the mass distribution we use the Kroupa (2002) stellar mass
function (MF) 
\begin{displaymath}
\frac{{\rm d} n_{1}}{{\rm d} M_{\star}} \propto \left\{
\begin{array}{ll}
M_{\star}^{-1.3} & \,\,\,\, 0.08 < M_{\star}/M_\odot < 0.5 \,, \\ 
M_{\star}^{-2.3} & \,\,\,\, 0.5 < M_{\star}/M_\odot < 50 \,, \\
\end{array} \right.
\end{displaymath}
which is a good fit to the local field population (see also Chabrier
2001).  
We ignore the
contribution from brown dwarfs as, due to the $M_{\star}^2$ factor, this
population - whilst numerous - makes only a small contribution
to the disruption rate. We normalise the mass function so that the total
mass density is $0.023 \, M_{\odot}\, {\rm pc}^{-3}$.
We take the speed distribution to be Gaussian about the mini-halo speed,
$V=270\, {\rm km \, s}^{-1}$,
\begin{equation}
\frac{{\rm d} n_{2}} {{\rm d} v} = \frac{1}{(2 \pi)^{1/2} \sigma_{\star}}
                   \exp{\left[- \frac{(v-V)^2}{2 \sigma_{\star}^2}
        \right]} \,,
\end{equation}
with stellar speed dispersion $\sigma_{\star}= 25 \, {\rm km \, s}^{-1}$.

The resulting one-off disruption timescales are $0.8 \, {\rm
Gyr}$ for halo 1 and $0.5 \, {\rm Gyr}$ for halo 2/3. 
For halo 2/3 the disruption time is similar to that
calculated assuming delta-function mass and speed distributions. The
main result though is that, once the spread in stellar masses is taken
into account, the more massive halo 1 can undergo one-off disruption
on a timescale smaller than the age of the MW. Taking into account the
spread of masses and velocities is therefore crucial for calculating
the mass threshold above which mini-halos will not lose a significant 
fraction of their mass in a single encounter.

\subsubsection{Disruption through multiple encounters}

The timescale on which a mini-halo will lose a significant fraction
of its mass as a result of the cumulative effects of
encounters with $\Delta E(b)/E<1$ (which, for compactness, we 
refer to a `disruption through multiple encounters') can
be estimated as
\begin{equation}
t_{\rm dis} = \frac{E}{({\rm d} E/ {\rm d} t)_{\rm tot}} \,.
\end{equation}
This is likely to be an overestimate; the mini-halo density profile
changes in response to interactions and this appears to 
reduce the effect of cumulative
interactions (Goerdt et al. 2006).

Starting, once again, with the simplifying assumption that all
stars have the same mass and relative velocity then
\begin{equation}
\frac{({\rm d} E/{\rm d} t)_{\rm tot}}{E} = 
 2 \pi \int_{b_{\rm c}}^{\infty} n v \frac{\Delta E(b)}{E} b \, {\rm d} b \,,
\end{equation}
and taking the same parameter values as above we find, for orbits
which lie entirely within the disc, $t_{\rm dis}=0.4 \, {\rm Gyr}$ 
for halo 1 and $t_{\rm dis}=0.5  \, {\rm Gyr}$ for halo 2/3.
The shorter timescale for multiple disruption for halo 1
reflects the fact that it can not undergo one-off disruption
and hence $b_{\rm c}=0$, whereas for halo 2/3 $b_{\rm c}=
0.0075 \, {\rm pc}$.

Generalising to a distribution of masses and relative velocities
the fractional energy input rate becomes
\begin{eqnarray}
&& \frac{({\rm d} E/{\rm d} t)_{\rm tot}}{E} = 
 2 \pi \int \int \nonumber \\ 
   &&
   \left[ \int_{b_{\rm c}(M_{\star}/v)}^{\infty} 
       \frac{{\rm d}^2 n}{{\rm d} M_{\star} {\rm d} v}
       v b \frac{\Delta E(b)}{E}  
         {\rm d} b \right] 
      {\rm d} v \, {\rm d} M_{\star} \,.
\end{eqnarray}
We now find $t_{\rm dis}= 0.6  \, {\rm Gyr}$ for halo 1 and
$t_{\rm dis} = 0.5 \,, {\rm Gyr}$ for haloes 2/3. 

\vspace{0.2cm}
The net energy input rate is the sum of the energy input rates
from 'one-off' and 'multiple disruption', and the net disruption timescale
will be shorter than the characteristic timescales for 'one-off'
and 'multiple disruption' individually.

\vspace{0.5cm}

We have assumed that the stellar density within the disc is uniform.
In general, clustering will increase the spread in disruption
timescales for mini-haloes of a given mass.  In addition most stars
(certainly those with $M_{\star}>0.5 M_\odot$) are in fact in binary
systems (e.g. Goodwin et al. 2006) and will cause a greater disruptive
effect than a single star.  Systems whose separations are
significantly less than the mini-halo radius ($< 1000$~AU) will
effectively combine the primary and secondary masses and, due to the
$M_\star^2$ dependence of the energy input, even fairly low-mass
secondaries may play an important role. We estimate that $\sim 30 -
40\%$ of stars with $M_{\star} > 1 M_\odot$ may have a large enough
companion to increase the energy input by a factor $>2$.\footnote{A
companion with a mass ratio greater than $0.4$ will increase the 
energy input by more than
$(1.4)^2 \sim 2$, and we assume a binary fraction of $\sim 60\%$ (see
Duquennoy \& Mayor 1991).} Even very low mass stars ($<0.5 M_\odot$)
have a binary frequency of $\sim 30\%$ (Fischer \& Marcy 1992), and so
the fraction of M-dwarfs with a companion that could very
significantly increase the energy input is $\sim 15 - 20\%$.
Thus, an accurate calculation of
mini-halo disruption will have to combine simulations of
mini-halo orbits in a realistic potential with an accurate model 
of the stellar distribution, including the binary fraction, within the disk.

\subsection{Mass dependence}

\begin{figure}
\setlength{\unitlength}{\linewidth}
\includegraphics[width=\linewidth]{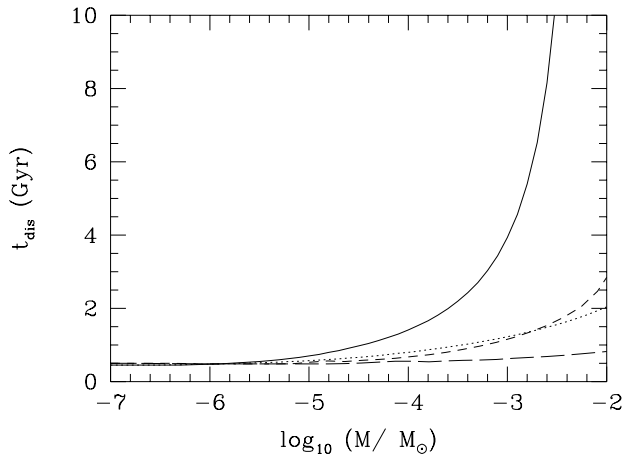}
\caption{The disruption timescales for an orbit entirely  within the disc 
as a function of mini-halo mass.
Solid line: one-off disruption assuming $b_{\rm s} \propto M^{0.33}$,
dotted: one-off disruption assuming $b_{\rm s} \propto M^{0.2}$,
short dashed: multiple disruption assuming $b_{\rm s} \propto M^{0.33}$,
long dashed: multiple disruption assuming $b_{\rm s} \propto M^{0.2}$.
}
\end{figure}

We have seen that more massive mini-haloes are less susceptible to
disruption.  It is therefore interesting to investigate the mass
dependence of the disruption timescales. Furthermore the WIMP
damping scale, 
and hence the mass of the smallest mini-haloes, depends on the 
properties (elastic scattering cross-section and mass) of the WIMPs. 
For generic WIMPs Green et al. (2005) found a spread in the minimum
mass of several orders of magnitude, while Profumo, Sigurdson
\& Kamionkowski (2006) have recently found that in the Minimal 
Supersymmetric Standard Model the minimum mass may vary between
$10^{-12} M_{\odot}$ and $10^{-4} M_{\odot}$.

The correct mass dependence of the small-$b$ energy input, and hence
the transition radius, 
$b_{\rm s}$, is not known. We consider two scalings which should give
an indication of the trend, and also the uncertainties. Firstly,
motivated by the the qualitative understanding of the mass dependence
of the asymptotic limits of the fractional energy input found in section
3.3, we once more consider a uniform density sphere.  For uniform
density spheres (with constant density) the profile 
parameter $A (=4\alpha^2/9\beta^2)$ is independent of the
radius/mass and hence $b_{\rm s} \propto R \propto M^{1/3}$.  For the
3 sample haloes $A$ decreases slightly with increasing mass 
(albeit with significant scatter between profiles and between haloes 2
and 3), and $b_{\rm s} \propto M^{0.2}$. We use both of these
scalings, normalising in both cases to $b_{\rm s}= 0.004 \, {\rm pc}$ at
$M=10^{-6} M_{\odot}$. The correct variation might be significantly
different from either of these scalings, however, and needs to be
determined from the profiles of simulated haloes with a range of
masses.  In Fig.~6 we plot the resulting disruption times for an orbit
entirely within the disc (for other orbits the disruption timescales
scale roughly as the fraction of time spent within the disc) as a
function of mini-halo mass, using the sudden transition approximation
with $(\Delta E/E)_{{\rm fid, s}}=10^{-8}$.

For very small mini-haloes, $M< 10^{-7} M_{\odot}$, the one-off and
multiple-encounter disruption timescales are independent of mass, and roughly equal. 
The mass independence is because for these small mini-haloes the transition 
impact parameter is smaller than the critical impact parameter for the
range of $M_{\star}$ and $v$ values considered, $b_{\rm s} <
b_{\rm c}(M_{\star}/v) $, so that $b_{\rm c}(M_{\star}/v)$ lies in the
$\Delta E(b)/E \propto b^{-4}$ regime and is independent of the
mini-halo mass. The approximate equality of the one-off and multi 
disruption timescales can be understood by considering the simplified case
of a delta-function mass/velocity distribution once more. Then, using
the sudden-transition approximation, both disruption timescales are
 equal to $[\pi n v (M_{\star}/M_{\odot})(300\, 
{\rm km \, s}^{-1}/v) (\Delta E/E)_{\rm fid, s}^{1/2} (1\, {\rm pc})^2]^{-1}$
if $b_{\rm s} < b_{\rm c}$.
The more rapid scaling of $b_{\rm s}$, 
with $M$ ($M^{0.33}$ versus $M^{0.2}$) also leads to larger values of
$t_{\rm dis}$.  The size of these differences increases with
increasing mini-halo mass. The exact disruption
timescales of more massive mini-haloes will depend 
on the mass dependence of the impact parameter at which the transition
between close and distant encounters occurs and also how rapidly this
transition occurs.

\subsection{Mini-halo radius}

In common with other studies (Zhao et al. 2005ab, Moore et al. 2005),
we have taken the mini-halo radii to be the radius at which the
density is 200 times the critical density at $z=26$ (the red-shift at
which Diemand et al. (2005) stopped their simulations and plotted the
profiles of their sample haloes), $r_{200}(z=26)$.  The densities of
simulated haloes do not decline sharply to zero beyond a given radius
however and, if the mini-haloes remained isolated beyond this
red-shift their nominal radii (and hence masses and binding energies)
would increase as the background density decreases.  As an extreme
example, if halo 1 remained isolated to z=0 then, assuming a NFW
density profile, its present day radius would be $\sim 0.7 \, {\rm
pc}$, its mass would a factor of $\sim 3$ larger
 and, using equations (\ref{deltaEa}) and (\ref{deltaEb}), the
fractional energy input in small (large) b encounters would be
substantially decreased (increased). The value used for the mini-halo
radius therefore has a potentially significant effect on calculations
of the fractional energy input.

Once a mini-halo is accreted onto a larger halo it no longer accretes
further mass onto itself and it is also subject to the tidal field of the
parent halo. For a mini-halo orbiting within a Milky Way-like parent
halo the tidal radius is only comparable to $r_{200}(z=26)$ for very
small, of order a few kpc, Galactocentric radii. The radius of a mini-halo 
which does not pass through the very central regions of the Milky Way will be 
the smaller of the tidal radius and the radius at the time of accretion 
(both of which are larger than $r_{200}(z=26)$). The red-shift at which
accretion occurs will, however, be different for different mini-halos
with the same initial properties. A detailed calculation of mini-halo
evolution will therefore have to include the mini-halo merger
histories. The majority of mini-halos, in particular those which pass
close to the solar radius and are hence most relevant for WIMP direct
and indirect searches, will be accreted onto larger halos not long
after z=26 and hence $r_{200}(z=26)$ should be a reasonable estimate
of their radii.

\section{Discussion}

We have studied the energy input into earth mass mini-haloes in
interactions with stars. Using the impulse approximation (see Spitzer
1958; Gerhard \& Fall 1983; Carr \& Sakellariadou 1999) we have
calculated the energy input as a function of impact parameter for a
range of mini-halo density profiles.  We also used the {\sc dragon}
code (e.g. Goodwin et al. 2004a,b; Hubber et al. 2006) to simulate
interactions with Plummer sphere haloes. We found excellent agreement
with the impulse approximation in the asymptotic limits $b \ll/\gg R$
(where $b$ is the impact parameter and $R$ is the mini-halo radius)
with a rapid transition at $b \sim 0.1 R$ between these regimes. We
also verified the scaling of the fractional energy input with stellar
mass and relative velocity.

Using analytic calculations we find that the fractional energy input
for large impact parameters, $b \gg R$, appears to be fairly
independent of the mini-halo mass, varying by a factor of $\sim 2$ for
haloes with masses which differ by a factor of $\sim 50$ with a
similar variation for different density profiles. This behaviour
probably reflects the fact that the haloes form at roughly the same
time and hence have similar characteristic densities. The fractional
energy
input in the $b \rightarrow 0$ limit depends quite strongly on the
mini-halo mass (being larger for lighter haloes) and is also dependent
on the central density profile. For the NFW profile, which has
asymptotic inner density profile $\rho \propto r^{-1}$, the fractional
energy input only becomes significantly larger than that for the cored
density profiles at tiny, and hence extremely rare, impact parameters
i.e. $b \ll 10^{-3} R$. This divergence is therefore essentially
unimportant for our calculations, however the central regions of
haloes with cuspy density profiles may be able to survive even after
substantial energy input/mass loss (e.g. Moore et al. 2005; Goerdt
et al. 2006). Motivated by
the results of our analytic and numerical calculations we formulate a
fitting function for the fractional energy input as a function of
impact parameter, which we refer to
as the `sudden transition' approximation. The slope of the fractional
energy input at large impact parameters is constant, while the impact
parameter which characterises the transition between the limits is
mini-halo mass dependent.

We also investigated the dependence of the critical impact parameter,
$b_{\rm c}$,
for which the energy input is larger than the mini-halo binding energy
on the mini-halo mass
and also the relative speed and mass of the interacting star. As
expected from the fractional energy input calculations, for slow
encounters with massive stars $b_{\rm c}$ is independent of the
mini-halo mass. There is a critical value of $(M_{\star}/v)$, which
increases with increasing halo mass, below which the energy input
is always smaller than the binding energy. For all values of
$(M_{\star}/M_{\odot})(300\, {\rm km \, s}^{-1}/v)$, the results of
our Plummer sphere simulations are in good agreement with the analytic
expressions for $b_{\rm c}$ from the sudden transition approximation.

We then use the sudden transition approximation to estimate the
timescales for one-off and multiple disruption for mini-haloes in the
MW as a function of mini-halo mass, using the approximate destruction 
criterion $\Delta E/ E = 1$. We take into account the stellar and velocity
distribution and note that binary stars can cause a significantly
greater energy input than single stars, due to their greater
effective mass.  For light mini-haloes with $M < {\cal O} (10^{-7}
M_{\odot})$ the disruption timescales are independent of mini-halo
mass and, for a mini-halo in the inner regions of the MW on a typical
orbit which spends a few per-cent of its time passing through the
disc, are comparable to the age of the MW. For more massive
mini-haloes, $M > {\cal O}(10^{-4} M_{\odot})$, the disruption
timescale estimates increase rapidly with increasing mass, suggesting
that the majority of these mini-haloes will not be disrupted by
stellar encounters. It is important to caution, however, that
the relationship between the energy input and the change in the bound mass
is not straight forward. In particular the mini-halo density profile
evolves so that successive multiple encounters are less effective than
would naively be expected and even if the energy
input in a single encounter
is much larger than the binding energy a small fraction of the mass
can remain bound (Goerdt et al. 2006).  Therefore these simple
estimates are likely to be overestimates of the actual disruption
timescales.

Finally we discussed the dependence of the fractional energy input on
the mini-halo radius assumed.  To be consistent with other studies
(Zhao et al. 2005ab, Moore et al. 2005), we took the radius to be the
radius at which the density is 200 times the critical density at
$z=26$, the red-shift at which Diemand et al. (2005) stopped their
simulations and plotted the profiles of their sample haloes.  This is
a somewhat arbitrary definition however; the densities of simulated
haloes do not decline to zero beyond this radius and as the background
density decreases the nominal radius increases. The physical
extent/radius will in fact be that at the time of accretion onto a
larger halo, or the tidal radius if this is smaller.  The tidal radius
within a Milky Way-like parent halo is only smaller than $r_{200}(z=26)$,
at small Galactocentric radii, however the majority of mini-halos will
be accreted onto larger halos shortly after this red-shift, so in
practice $r_{200}(z=26)$ should be a reasonable estimate of the radius
of most mini-halos.

A complete calculation of the disruption of mini-halos will need to
take into account their merger histories, simultaneously and
consistently incorporate disruption due to encounters with stars and
tidal stripping. Mini-halos formed from rare, large density
fluctuations, will be denser, and hence more resilient to disruption,
than typical mini-halos and this will also need to be included.

\section*{Acknowledgements}
AMG is supported by PPARC and SPG by the UK Astrophysical Fluids
Facility (UKAFF).  We are grateful to Tobias Goerdt and Simon White 
for useful comments/discussion.

\label{lastpage}


\begin{thebibliography}{}

\bibitem{ahw} Aarseth, S., H\'enon, M. \& Wielan, R., 1974, A\&A, 37, 183

\bibitem{aw} Aguilar, L. A., \& White, S. D. M., 1985, ApJ, 295, 374

\bibitem{az} Angus, G. W., \& Zhao, H., 2006 [astro-ph/0608580]

\bibitem{bh} Barnes, J. \& Hut, P., 1986, Nat, 324, 446

\bibitem{bde} Berezinsky, V., Dokuchaev V. \& Eroshenko. Y., 2003,
   Phys. Rev. D 68, 103003

\bibitem{bde2} Berezinsky, V., Dokuchaev V. \& Eroshenko. Y., 2006,
   Phys. Rev. D 73, 063504

\bibitem{bGZ} Berezinsky, V., Gurevich A. V. \& Zybin, K. P.,
       1992, Phys. Lett. B 294, 221

\bibitem{bhs} Bertone, G., Hooper, D., \& Silk, J., 2005, Phys. Rept. 405,
           279

\bibitem{cs} Carr, B. J. \&  Sakellariadou, M., 1999, ApJ, 516, 195

\bibitem{c} Chabrier, G., 2001, ApJ, 554, 1274


\bibitem{first} Diemand, J., Moore, B. \& Stadel, J., 2005, Nat, 433, 
          389  

\bibitem{dm91} Duquennoy, A. \& Mayor, M., 1991, A\&A, 248, 485

\bibitem {fm92} Fischer, D. A. \& Marcy, G. W., 1992, ApJ, 396, 178

\bibitem{gf} Gerhard, O. E. \& Fall, S. M., 1983, MNRAS, 203, 1253


\bibitem{go} Gnedin, O. Y. \& Ostriker, J. P., 1999, ApJ, 513, 626

\bibitem{ggmds} Goerdt, T. et al., 2006 [astro-ph/0608495]

\bibitem{g97} Goodwin, S. P. 1997, MNRAS, 284, 785

\bibitem{wwt04a} Goodwin, S. P., Whitworth, A. P. \& Ward-Thompson,
  D. 2004a, A\&A, 414, 633

\bibitem{gwwt04b} Goodwin, S. P., Whitworth, A. P. \& Ward-Thompson,
  D. 2004b, A\&A, 423, 169

\bibitem{gkgb06} Goodwin, S. P., Kroupa, P., Goodman, A. \& Burkert, A.,
  2006, to appear in 'Protostars and Planets V', [astro-ph/0603233]



\bibitem{ghs} Green, A. M., Hofmann, S. \& Schwarz, D. J., 2004,
             MNRAS, 353 L23

\bibitem{ghs2} Green, A. M., Hofmann, S. \& Schwarz, D. J., 2005,
              JCAP, 08, 003




\bibitem{hss} Hofmann, S., Schwarz, D. J. \& St\"ocker, H., 2001,
         Phys. Rev. D 64, 083507

\bibitem{hubber} Hubber, D. A., Goodwin, S. P. \& Whitworth, A. P., 2006,
  A\&A, 450 881

\bibitem[]{}Kazantzidis, S., Magorrian, J. \& Moore, B. 2004, ApJ,
  601, 37



\bibitem{kk} Klypin, A., Kravtsov, A. V. 
       Valenzuela, O. \& Prada, F., 1999, ApJ, 522, 82



\bibitem{k2} Kroupa, P., 2002, Sci, 295, 82


\bibitem{kg} Kuijken, K. \& Gilmore, G., 1989, MNRAS 289, 605

\bibitem{lz} Loeb, A. \& Zaldarriaga, M., 2005, Phys. Rev. D 71, 103520 

\bibitem{mooreimp} Moore, B., 1993, ApJ, 413, L93 

\bibitem{m} Moore, B. et al., 1999, ApJ, 524, L19

\bibitem{mooreresp} Moore, B., Diemand, J., Stadel, J. \& Quinn, T., 2005,
[astro-ph/0502213]

\bibitem{nfw1} Navarro, J. F., Frenk, C. S. \& White, S. D. M., 1996,
           ApJ, 462, 563

\bibitem{nfw2} Navarro, J. F., Frenk, C. S. \& White, S. D. M., 1997,
           ApJ, 490, 493


\bibitem{ol} Oguri, M. \& Lee, J., 2004, MNRAS, 355 120 

\bibitem{pb}  Pe\~narrubia, J. \& Benson, A. J., 2005, MNRAS 364, 977

\bibitem{plummer} Plummer, H. C., 1915, MNRAS 76, 107 




\bibitem{psk} Profumo, S., Sigurdson, K. \& Kamionkowski, M.,
          2006, Phys. Rev. Lett. 97 031301




\bibitem{s} Spitzer, L. Jr.,1958,  ApJ, 127, 17

\bibitem{shs} Schwarz, D. J., Hofmann, S. \& St\"ocker, H., 2001,
              PrHEP hep2001 204 [astro-ph/0110601]


\bibitem{tb} Taylor, J. E. \& Babul, A., 2004, MNRAS 348, 811





\bibitem{zb} Zenter, A. \& Bullock, J., 2003, MNRAS 348, 811

\bibitem{zhao} Zhao H., Taylor, J.  E., Silk, J. \& Hooper, D., 2005,  
[astro-ph/0502049]

\bibitem{zhao2} Zhao H., Taylor, J.  E., Silk, J. \& Hooper, D., 2005b, 
[astro-ph/0508215]

\end{thebibliography}
\end{document}